\documentclass[conference]{IEEEtran}
\usepackage{simplemargins}
\setleftmargin{16mm} \setrightmargin{13mm}
\settopmargin{19mm}  \setbottommargin{43mm}% ---------------
\makeatletter

\def\ps@headings{%
\def\@oddhead{\mbox{}\scriptsize\rightmark \hfil \thepage}%
\def\@evenhead{\scriptsize\thepage \hfil \leftmark\mbox{}}%
\def\@oddfoot{}%
\def\@evenfoot{}}
\makeatother
\usepackage{graphicx}
\usepackage{indentfirst}
\usepackage{epsfig}
\usepackage{amsfonts}
\usepackage{color}
\usepackage{amsmath}
\title{Control and Optimization Meet the Smart \\Power Grid: Scheduling of Power Demands \\for Optimal Energy Management}
\author{\large{Iordanis Koutsopoulos} \qquad \large{Leandros Tassiulas} \\
Department of Computer and Communications Engineering, University of Thessaly \\
and Center for Research and Technology Hellas (CERTH), Greece\\Emails: \{leandros,jordan\}@uth.gr}
\begin{document} \maketitle

\newtheorem{property}{Property}
\newcommand{\be}{\begin{itemize}} \newcommand{\ee}{\end{itemize}}
\newcommand{\tb}{\textbf} \newcommand{\ttt}{\texttt}
\newcommand{\tit}{\textit} \newcommand{\uline}{\underline}
\newtheorem{proposition}{Proposition}
\newtheorem{conjecture}{Conjecture}
\newtheorem{theorem}{Theorem} \newtheorem{lemma}{Lemma}
\newtheorem{fact}{Fact}
\newtheorem{rem}{Remark}

\begin{abstract}
The smart power grid aims at harnessing information and communication technologies to enhance reliability and enforce sensible use of energy. Its realization is geared by the fundamental goal of effective management of demand load. In this work, we envision a scenario with real-time communication between the operator and consumers. The grid operator controller receives requests for power demands from consumers, each with different power requirement, duration, and a deadline by which it is to be completed. The objective of the operator is to devise a power demand task scheduling policy that minimizes the grid operational cost over a time horizon. The operational cost is a convex function of instantaneous total power consumption and reflects the fact that each additional unit of power needed to serve demands is more expensive as the demand load increases.

First, we study the off-line demand scheduling problem, where parameters are fixed and known a priori. If demands may be scheduled preemptively, the problem is a load balancing one, and we present an iterative algorithm that optimally solves it. If demands need to be scheduled non-preemptively, the problem is a bin packing one. Next, we devise a stochastic model for the case when demands are generated continually and scheduling decisions are taken online and focus on long-term average cost. We present two instances of power consumption control based on observing current consumption. In the first one, the controller may choose to serve a new demand request upon arrival or to postpone it to the end of its deadline. The second one has the additional option to activate one of the postponed demands when an active demand terminates. For both instances, the optimal policies are threshold-based. We derive a lower performance bound over all policies, which is asymptotically tight as deadlines increase. We propose the Controlled Release threshold policy and prove it is asymptotically optimal. The policy activates a new demand request if the current power consumption is less than a threshold, otherwise it is queued. Queued demands are scheduled when their deadline expires or when the consumption drops below the threshold.
\end{abstract}

\section{Introduction}

The smart power grid is currently considered a major challenge for harnessing information and communication technologies to enhance the electric grid flexibility and reliability, enforce sensible use of energy and enable embedding of different types of grid resources to the system. These resources include renewable ones, distributed micro-generator customer entities, electric storage, and plug-in electric vehicles \cite{moslehi}. The smart power grid shall incorporate new technologies that currently experience rapid progress, such as advanced metering, automation, bi-directional communication, distributed power generation and storage. The ultimate interconnection and real-time communication between the consumer and the market/system operator premises will be realized through IP addressable components over the internet \cite{lui}.

The design and realization of the smart power grid is geared by the fundamental goal of effective management of power supply and demand loads. Load management is primarily employed by the power utility system operator with the objective to match the power supply and demand profiles in the system. Since the supply profile shaping depends highly on demand profile, the latter constitutes the primary substrate at which control should be exercised by the operator. The basic objective therein is to \tit{alleviate peak load by transferring non-emergency power demands at off-peak-load time intervals}.

Demand load management does not significantly reduce total energy consumption since most of the curtailed demand jobs are transferred from peak to off-peak time intervals. Nevertheless, load management aids in \tit{smoothing} the power demand profile of the system across time by avoiding power overload periods. By continuously striving to maintain the total demand to be satisfied below a critical load, grid reliability is increased as grid instabilities caused by voltage fluctuations are reduced. Further, the possibility of power outage due to sudden increase of demand or contingent malfunction of some part of the system is decreased. More importantly, demand load management reduces or eliminates the need for inducing supplementary generated power into the grid to satisfy increased demand during peak hours. This supplementary power is usually much more costly to provide for the operator than the power for average base consumed load, since it emanates from gas micro-turbines or is imported from other countries at a high price. Thus, from the point of view of system operator, \tit{effective demand load management reduces the cost} of operating the grid, while from the point of view of the user, it lowers real-time electricity prices.

In this paper, we make a first attempt to formulate and solve the basic control and optimization problem faced by the power grid operator so as to achieve the goals above. We envision a scenario with real-time communication between the operator and consumers through IP addressable smart metering devices installed at the consumer and operator sides. The grid operator has full control over consumer appliances. The operator controller receives power demand requests from different consumers, each with different power requirements, different duration (which sometimes may be even unknown), and different flexibility in its satisfaction. Flexibility is modeled as a deadline by which each demand needs to be completed. The objective of the grid operator is to \tit{devise a power demand task scheduling policy that minimizes the grid operational cost} over a time horizon. The operational cost is modeled as a convex function of instantaneous total power consumption in the system, so as to reflect the fact that each additional Watt of power needed to serve power demands is more expensive as the total power demand increases.

\subsection{State-of-the-art}

In the power engineering terminology, the power demand management method above is known as \tit{demand response} support \cite{link1}. Demand response is currently realized mostly through static contracts that offer consumers lower prices for the power consumed at off-peak hours, and they rely on customer voluntary participation. A recent development involves real-time pricing but still needs manual turning off of appliances. Currently, there exists significant research activity in automating the process of demand response through developing appropriate enabling technologies that reduce power consumption at times of peak demand \cite{hamilton}. GridWise \cite{link2} is an important research initiative in USA with this goal.

In one form, the automation process may involve regulation of power consumption level of consumer appliances like heaters or air conditioners (A/C) by the operator, or slight delaying of consumption until the peak demand is reduced. For instance, in the Toronto PeakSaver AC pilot program \cite{link3}, the operator can automatically control A/Cs during peak demand through an installed switch at the central A/C unit, thus in essence shifting portions of power consumption in time. Lockheed Martin has developed the \tit{SeeLoad}{\small\texttrademark} system \cite{link4} to realize efficient demand response in real-time. Other efforts like the \tit{EnviroGrid}{\small\texttrademark} by REGEN Energy Inc. are based on self-regulation of energy consumption of appliances within the same facility without intervention of the operator, through controllers connected in a ZigBee wireless network \cite{link5}. In an automated dynamic pricing and appliance response scenario, the work \cite{li} addresses the decision problem faced by home appliances of when to request instantaneous power price from the grid so as to perform power consumption adaptation. The problem is modeled as a Markov Decision Process subject to a cost of obtaining the price information.

At the level of modeling abstraction, the problem of smoothing power demand bears may slightly relate to that of scheduling tasks under deadline constraints in order to optimize total cost over a time horizon. There exists much literature on machine scheduling under deadline constraints in operations research literature, for optimizing mainly linear functions of the load \cite[Chap.21-22]{handbook}. For wire-line networks, the Earliest Deadline First (EDF) scheduling rule is optimal in the sense of minimizing packet loss due to deadline expirations \cite{panwar}.

Scheduling under deadlines with convex cost models gained momentum recently in wireless networks because the expended transmission energy is convex in throughput. In \cite{keslassy}, the authors solve the deterministic scheduling problem with a priori known packet arrival times under certain deadlines so as to minimize the total consumed energy if only one packet can be transmitted at a time. The work in \cite{neely} studies properties of the optimal off-line solution for the same problem and proposes heuristic online scheduling algorithms. In \cite{fu1} the authors consider the problem of minimizing the energy needed to transmit a certain amount of data within a given time interval over a time-varying wireless link. Energy is a convex function of the amount of data sent each time. The non-causal problem when link quality is known a priori is solved by convex optimization. The online problem where link quality each time is revealed to the controller just before decision is solved by dynamic programming. The optimal policy is of threshold type on the energy cost of sending data immediately versus saving it for later transmission. The multi-user version of the problem is studied in \cite{tarello}. In \cite{zafer} the problem of transmit rate control for minimizing energy over a finite horizon is solved through continuous time optimization, and optimal transmission policies are derived in terms of continuous functions of time that satisfy certain quality of service curves. Finally, the works \cite{kumar1}, \cite{kumar2} present a long-term view on probabilistic latency guarantees per packet in wireless networks based on a primal-dual-like algorithm for a utility maximization problem with a constraint on latency guarantees.

\subsection{Our contribution}

In this paper, we address the problem of optimal power demand scheduling subject to deadlines in order to minimize the cost over a time horizon. The problem is faced by a grid operator that has full control over the consumer appliances. To the best of our knowledge, this is the first work that attempts to characterize structural properties of the problem and the solutions in the context of smart grid power demand load management.
%The problem is fundamentally different from the ones above that have been addressed in the %context of wireless networks. The main difference is that the cost objective is on the total %power consumption which changes based on the arrival pattern, durations and power requirements %of tasks and our control policy.
The contribution of our work to the literature is as follows:
\be
\item We formulate the \tit{off-line version of the demand scheduling problem} for a certain time horizon, where the demand task generation pattern, duration, power requirement and deadline for each task are fixed and given a priori. We distinguish between elastic and inelastic demands that give rise to preemptive and non-preemptive task scheduling respectively. In the first case, the problem is a load balancing one, and we present an iterative algorithm that optimally solves it. In the second case, the problem is equivalent to bin packing, and thus it is NP-Hard.
\begin{figure}[htb]
\begin{center}
\epsfig{figure = 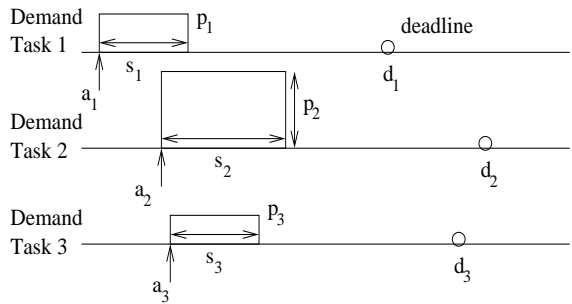, width=7.5cm,height=4cm}
\end{center}
\caption{Power demand task related parameters. Power demand task $n=1, 2, 3$ is generated at time $a_n$, has duration $s_n$, power requirement $p_n$ and needs to be completed by $d_n$.} \label{fig:tasks}
\end{figure}

\begin{figure}[htb]
\begin{center}
\epsfig{figure = 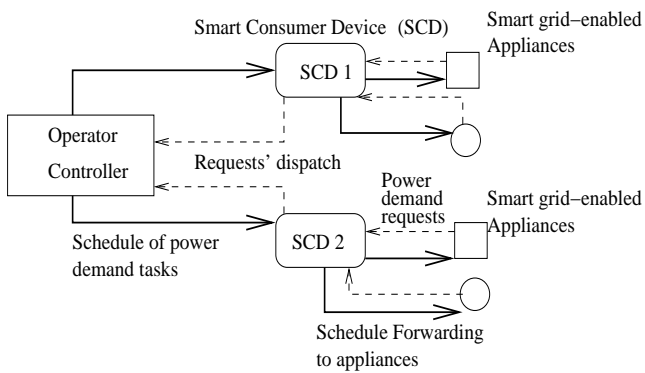, width = 8.6cm,height=4.9cm}
\end{center}
\caption{Overview of system architecture. The smart grid-enabled appliances send power demand requests to the smart consumer device, which further dispatches them to the controller at the operator side. The controller returns a schedule for each task which is passed to the appliances through the smart consumer device.} \label{fig:architecture}
\end{figure}
\item We study the \tit{online dynamic scheduling problem}. We propose a stochastic model for the case when demands are generated continually and scheduling decisions are taken online, and we consider minimizing the long-term average cost. First, we derive the performance of the simplest default policy which is to schedule each task upon arrival. Next, we present two instances of power consumption control based on observing current power consumption. In the first one, the controller may choose to serve a new demand request upon arrival or to postpone it until the end of its deadline. The second one is more enhanced and has the additional option to activate one of the postponed demands when a demand under service terminates. For both instances above, the optimal policies are threshold-based.
\item We derive a lower performance bound over all policies, which is asymptotically tight as deadlines increase by showing a sequence of policies that achieves the bound. We propose the Controlled Release threshold policy and prove it is \tit{asymptotically optimal} in that it achieves the bound specified above. The policy activates a new demand request if the current power consumption is less than a threshold, otherwise it is queued. Queued demands are scheduled when their deadline expires or when the consumption drops below the threshold.
\ee
The paper is organized as follows. In section \ref{sec:2} we present the model and assumptions and in section \ref{sec:3} we study the off-line version of the problem. Section \ref{sec:4} contains the study about the online version of the problem, the derived lower bound and the optimal policies, and section \ref{sec:5} concludes our study.

\section{The Model} \label{sec:2}

We consider a controller located at the electric utility operator premises, with bi-directional communication to some smart devices each of which is located at a consumer's premises. Each smart device at a consumer side is connected to smart grid enabled appliances. The smart device collects power demand requests from individual appliances. These requests can be either manually entered by the user at the times of interest or they can be generated based on some automated process. Each power demand task $n$, $n=1,2,\ldots,$ has a time of generation $a_n$, a time duration $s_n$ time units, and an instantaneous power requirement $p_n$ (in Watts) when the corresponding task is activated and consumes power. Each task is characterized by some temporal flexibility or \tit{delay tolerance} in being activated, which is captured by a deadline $d_n \geq a_n$ by which it needs to be completed. For example, some appliances (e.g. lights) have zero delay tolerance, while others (e.g. washing machine) have some delay tolerance. Figure \ref{fig:tasks} depicts the parameters defined above for three tasks.
%We fix our attention to a finite horizon $T > \max_i d_i$.

We assume that all demand tasks shall be eventually scheduled, at the latest by their deadlines. In other words, there are no demand task losses in the system. A task may be scheduled to take place \tit{non-preemptively} or \tit{preemptively}. In the first case, once it starts, a task $n$ is active for $s_n$ \tit{consecutive} time units until completion. Thus, each task is scheduled at some time $t_n \in [a_n, d_n-s_n]$, or in other words it is scheduled with a time shift $\tau_n \in [0, D_n]$ after its arrival, where $D_n = d_n-s_n-a_n$. In the case of preemptive scheduling, each task $n$ may be scheduled with interruptions within the prescribed tolerance interval as long as it is finished on time. We assume that the instantaneous power consumption $p_n$ of a task cannot be adapted by the controller. Nevertheless, the possibility of having adaptable $p_n$ by the operator controller could be incorporated in our formulation.

The controller receives power demand requests from smart devices and it needs to decide on the time that the different power demand tasks are activated. Then, it sends the corresponding command for activation to the smart device from which the task emanated. The smart device transfers the command to the corresponding appliance, and the power demand is activated at the time prescribed by the operator controller (Fig. \ref{fig:architecture}). The communications from the controller to the smart devices and from them to the appliances take place through a high-speed connection and thus incur zero delay. We assume that the the grid operator has full control over the individual consumer appliances and that the appliances comply to the dictated schedule and start the task at the prescribed time.

\begin{figure}[htb]
\begin{center}
\epsfig{figure = 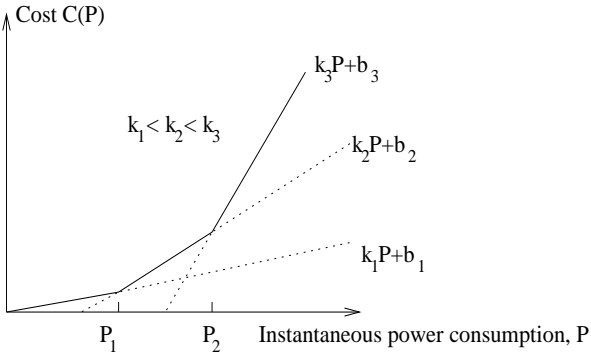, width=8.2cm,height=4.7cm}
\end{center}
\caption{A piecewise linear convex cost function $C(P)$ of instantaneous power consumption $P$ with $L=3$ power consumption classes, defined by lines $k_iP + b_i$, $i=1,2,3$. Two cross-over points $P_1, P_2$ distinguish the different classes.} \label{fig:convex}
\end{figure}

In this work, we consider two versions of the problem:
\begin{enumerate}
\item An \tit{off-line} one for cost minimization over a time horizon, where the power demand generation times, durations, power requirements and deadlines are known non-causally to the controller. This is valid for cases where off-line scheduling can be used. Under those non-realistic assumptions we also obtain performance bounds.
\item An \tit{online} one for long-term average cost minimization, where quantities are stochastic. This fluid model captures the case where demands are generated continually and scheduling decisions are taken online.
\end{enumerate}

\subsection{Cost Model}

At each time $t$, let $P(t)$ denote the total instantaneous consumed power in the system. This is the summation over all active tasks, i.e. tasks that consume power at time $t$. We denote instantaneous cost associated with power consumption $P(t)$ at time $t$ as $C(P(t))$, where $C(\cdot)$ is an increasing, differentiable convex function. Convexity of $C(\cdot)$ reflects the fact that the differential cost of power consumption for the electric utility operator increases as the demand increases. That is, each unit of additional power needed to satisfy increasing demand becomes more expensive to obtain and make available to the consumer. For instance, supplementary power for serving periods of high demands may be generated from expensive means, or it may be imported at high prices from other countries. In its simplest form, the cost may be a piecewise linear function of the form:
\begin{equation}
C(x) = \max_{i=1,\ldots,L} \{k_i x + b_i\}
\end{equation}
with $k_1 \leq \ldots \leq k_L$, accounting for $L$ different classes of power consumption, where each additional Watt consumed costs more when at class $\ell$ than at class $(\ell-1)$, $\ell=2,\ldots,L$ (Fig. \ref{fig:convex}). In our model, we shall consider a generic convex function $C(\cdot)$.

\section{The Off-line Demand Scheduling Problem} \label{sec:3}

First, we consider the off-line version of the demand scheduling problem for $N$ power demand tasks. For each task $n=1,\ldots,N$, the generation time $a_n$, power consumption $p_n$, duration $s_n$ and deadline $d_n$ are deterministic quantities which are non-causally known to the controller before time $t=0$. This version of the problem may arise if task properties can be completely predictable (for instance, if tasks exhibit time periodicity) and in any case provides useful performance bounds. Fix attention to a finite horizon $T$.

\subsubsection{Preemptive scheduling of tasks}

Consider first the case of \tit{elastic} demands, which implies that each demand task $n$ may get \tit{preemptive} service, i.e. it does not need to be served contiguously. Namely, each task may be interrupted and continued later such that it is active at nonconsecutive time intervals, provided of course that it will be completed by its specified time $d_n$.
%We present the discrete-time version of the problem; the continuous-time case could be formulated similarly.
Each task $n$ has fixed power requirement $p_n$ when it is active.

For each task $n$ and time $t$, define the function $x_n(t)$, which is $1$, if job $n$ is active at time $t$, $t \in [0,T]$, and $0$ otherwise. A scheduling policy is a collection of functions $\mathbf{X} = \{x_1(t),\ldots,x_N(t)\}$, defined on interval $[0,T]$. The controller needs to find the scheduling policy that minimizes the total cost in horizon $[0,T]$, where at each time $t$, the instantaneous cost is a convex function of total instantaneous power load. The optimization problem faced by the controller is:
\begin{equation}
\min_{\mathbf{X}} \int_0^T \! C\bigg(\sum_{n=1}^N p_n x_n(t)\bigg)\, dt
\label{eq:problem}
\end{equation}
subject to:
\begin{equation}
\int_{a_n}^{d_n} \! x_n(t)\,dt = s_n\,.
\label{eq:con}
\end{equation}
and $x_n(t) \in \{0,1\}$ for all $n=1,\ldots,N$ and $t \in [0,T]$. The constraint implies that each task should be completed by its respective deadline.

The problem above is combinatorial in nature due to binary-valued functions $x_n(t)$. A lower bound in the optimal cost is obtained if we relax $x_n(t)$ to be continuous-valued functions, so that $0 \leq x_n(t) \leq 1$. This relaxation allows us to capture the scenario of varying instantaneous power level for each task $n$; at time $t$, $p_n x_n(t)$ denotes the instantaneous consumed power by demand task $n$. For each $n=1,\ldots,N$, define the set of functions that satisfy feasibility condition (\ref{eq:con}),
\begin{equation}
\mathcal{F}_n = \{x_n(t) : \int_{a_n}^{d_n} \! x_n(t)\,dt = s_n\}
\end{equation}
with $0 \leq x_n(t) \leq 1$ for all $t \in [0,T]$.

The following fluid model captures the continuous-valued problem. Consider the following bipartite graph $\mathcal{U} \cup \mathcal{V}$. There exist $|\mathcal{U}| = N$ nodes on one side of the graph, one node for each task. Also, there exist $|\mathcal{V}|$ nodes, where each node $k$ corresponds to the infinitesimal time interval $[(k-1)\,dt, k\,dt]$ of length $dt$. From each node $n=1,\ldots,|\mathcal{U}|$, we draw links towards infinitesimal time intervals that reside in interval $[a_n,d_n]$. Input flow $p_n s_n$ enters each node $n=1,\ldots, |\mathcal{U}|$. Let $\ell(t) =\sum_{n=1}^N p_n x_n(t)$ denote the power load at time $t$, $0 \leq t \leq T$.

The problem belongs to the class of problems that involve the sum (here, integral) of convex costs of loads at different locations (here, infinitesimal time intervals),
\begin{equation}
\min \int_0^T \! C\big(\ell(t)\big)\,dt\,,
\end{equation}
and for which the solution is load balancing across different locations \cite{hajek}.

For given load function $\ell(t)$, define the operator $\mathcal{T}_n$ on $\ell(t)$ as:
\begin{equation}
\mathcal{T}_n \ell(t) = \arg\min_{x_n(t) \in \mathcal{F}_n} \int_0^T \! C\big(\ell(t)\big)\,dt\,.
\label{eq:operator}
\end{equation}
Now define a sequence of demand task indices $\{i_k\}_{k \geq 1}$ in which tasks are parsed. One such sequence is $\{1,\ldots,n,1,\ldots,n, \ldots\}$, where tasks are parsed one after the other according to their index in successive rounds. Consider the sequence of power load functions $\ell^{(k+1)}(t) = \mathcal{T}_{i_k} \ell^{(k)}(t)$, for $k=1,2,\ldots$. For example, if $i_k = n$, the problem
\begin{equation}
\min_{x_n(t) \in \mathcal{F}_n} \int_0^T \!C\big(p_n x_n(t) + \sum_{k \neq n} p_k x_k(t)\big)
\end{equation}
with $0 \leq x_n(t) \leq 1$, $0 \leq t \leq T$, is solved in terms of $x_n(t)$, while other functions $x_k(t)$, $k \neq n$ are kept unchanged. This is a convex optimization problem, for which the KKT conditions yield the solution function $x_n(t)$. Essentially this is the function that balances power load across times $t \in [0,T]$ as much as possible at that iteration.

\begin{theorem}
The iterative load balancing algorithm that generates the sequence of power load functions $\ell^{(k+1)}(t) = \mathcal{T}_{i_k} \ell^{(k)}(t)$, for $k=1,2,\ldots$, where operator $\mathcal{T}$ is defined by (\ref{eq:operator}) converges to the optimal solution for the continuous-valued problem (\ref{eq:problem})-(\ref{eq:con}).
\end{theorem}

\begin{proof}
In \cite[pp.1403-1404]{hajek} a proof methodology is developed for the case of discrete locations and discrete flow vectors. It is straightforward to extend this methodology to the instance described here, with the integral in the objective and functions $x_n(t)$ instead of the discrete vectors, to show that the sequence of power load functions $\ell^{(k)}(t)$, for $k=1,2,\ldots$ converges to the optimal solution $\mathbf{X}^*$ for the original continuous-valued problem and that the final optimal set of functions, $\mathbf{X}^*$ minimizes the maximum power load over all times $t$. The corresponding problem with binary-valued functions $\{x_n(t)\}$ has similar properties, as discussed in \cite{hajek}.
\end{proof}

\subsubsection{Non-preemptive scheduling of tasks}

Now, we consider the case of \tit{inelastic} demands. Namely, we assume that, once scheduled to start, a task should be served uninterruptedly until completion. A discrete-time consideration is better suited to capture this case. Consider the following instance $\mathcal{I}$ of the problem. For each task $n=1,\ldots,N$, let the generation time $a_n = 0$ and the deadline $d_n =D$, i.e. common for all tasks. Also assume that power requirements are the same, i.e. $p_n = p$ for all $n$. Fix a positive integer $m$, and consider the following \tit{decision version} of the scheduling problem: Does there exist a schedule for the $N$ tasks such that the maximum instantaneous consumed power is $mp$?

Let us view each task $n$ of duration $s_n$ as an item of size $s_n$, and the horizon $T = D$ as a bin of capacity $D$. Then, the question above can be readily seen to be equivalent to the decision version of the \tit{one-dimensional bin packing} problem: ``Does there exist a partition of the set of $N$ items into $m$ disjoint subsets (bins) $U_1,\ldots,U_m$, such that the sum of the sizes of items in each subset (bin) is $D$ or less?'' Clearly, each bin is one level of step $p$ of power consumption. If $m$ bins suffice to accommodate the $N$ items, then the maximum instantaneous power consumption is $mp$, and vice versa.

The optimization version of the one-dimensional bin packing problem is to partition the set of $N$ items into the \tit{smallest} possible number $m$ of disjoint subsets (bins) $U_1,\ldots, U_m$ such that the sum of the sizes of items in each subset (bin) is $D$ or less. This is equivalent to the problem of finding a schedule of power demand tasks that minimizes the maximum power consumption over the time horizon $T$. Minimizing the maximum power consumption in the time horizon of duration $T$ was shown to be equivalent to minimizing the total convex cost in the horizon. The decision version of bin packing is NP-Complete \cite{garey}, and thus the optimization version of bin packing is NP-Hard. It can thus be concluded that \tit{finding a schedule that minimizes the total convex cost in the horizon is an NP-Hard problem.}

For different generation times $a_n$ and deadlines $d_n$, one can easily create instances that are equivalent to the bin packing problem. For different power requirements $p_n$, one way to proceed is to show equivalence with bin packing by defining a minimum quantum $\Delta_p$ of power requirements and by observing that a task with power requirement $p_n = n\Delta_p$ and duration $s_n$ is equivalent to $n$ tasks of size $\Delta_p$ and duration $s_n$.

\section{The Online Dynamic Demand Scheduling Problem} \label{sec:4}

We now consider the \tit{online dynamic version} of the scheduling problem. This captures the scenario where demands are generated continually and scheduling decisions need to be taken \tit{online} as the system evolves. Power demand requests arrive at the grid operator controller according to a Poisson process, with average rate $\lambda$ requests per unit of time. The time duration $s_n$ of each power demand request $n=1,2,\ldots$ is a random variable that is exponentially distributed with parameter $s$, i.e $\Pr(s_n \leq x) = 1 - e^{-sx}$, with $x \geq 0$. Equivalently, the mean request duration is $1/s$ time units, and $s$ is the average service rate for power demand tasks. The durations of different requests are independent random variables.

The deadline $d_n$ of each request $n=1,2,\ldots$ is also exponentially distributed with parameter $d$, i.e. $\Pr(d_n \leq x) = 1 - e^{-dx}$, with $x \geq 0$. Thus, the mean deadline is $1/d$ time units, and $d$ may be viewed as the \tit{deadline expiration rate}. Deadlines of different requests are independent.

We are interested in \tit{minimizing the long-run average cost} %$\mathbb{E}[C\big(P(t)\big)]$,
\begin{equation}
\lim_{T \rightarrow +\infty} \frac{1}{T} \mathbb{E} \big[\int_0^T \! C\big(P(t)\big) \,dt \big] = \mathbb{E}[C\big(P(t)\big)]\,,
\end{equation}
where the expectation is with respect to the stationary distribution of $P(t)$. A remark is in place here about the nature of system state that is assumed to be available to the grid operator controller. The controller can measure total instantaneous power consumption. This is a readily available type of state and a basic one on which control decisions should rely. There also exist other evolving parameters that could enhance system state, but we refrain from using these for decision making in this paper, mainly because our primary objective is on understanding the structure of simple control policies first before proceeding to more composite ones.

\subsection{Default Policy: No scheduling}

Consider the default, naive policy where each power demand is activated by the controller immediately upon its generation, namely there is no scheduling regulation of demand tasks. This policy is oblivious to instantaneous power consumption $P(t)$ and all other system parameters.

\subsubsection{Fixed power requirement per task}

First, assume that the power requirement of each task is fixed and unit, i.e. $p_n = 1$. The instantaneous power consumption at time $t$ is $P(t) = N(t)$, where $N(t)$ is the number of active demands at time $t$. Under the assumptions stated above on the demand arrival and service processes, $N(t)$ (and thus $P(t)$) is a continuous-time Markov chain. In fact, since each power demand task is always activated (served) upon arrival and there is no waiting time or loss, we can view $P(t)$ as \tit{the occupation process of an $M/M/\infty$ service system}.
%We can discretize the Markov chain by dividing time into small time windows of length $\Delta$. %The states of the discrete time Markov chain are $N_k = N(k \Delta)$, namely the number of active power demands at time $k\Delta$, $k=0,1,\ldots$.
From state $P(t)$, there are transitions to state:
\begin{itemize}
\item $P(t)+1$ with rate $\lambda$, when new demand requests arrive.
\item $P(t)-1$ with rate $P(t)s$, when one of the current $P(t)$ active demands is completed.
\end{itemize}
Through steady-state probabilities $q_i = \lim_{t \rightarrow \infty} \Pr(P(t)=i)$, $i=1,2,\ldots,$ and equilibrium equations we can obtain the steady-state probability distribution of the number of active power demand tasks,
\begin{equation}
q_i = {\left(\frac{\lambda}{s}\right)}^i \,\frac{e^{-\lambda / s}}{i\, !}\,,
\end{equation}
which is Poisson distributed with parameter $\lambda/s$. The same steady-state distribution emerges for an $M/G/\infty$ queue \cite{bertsekas}. Thus, the expected number of active requests in steady steady is simply $\mathbb{E}[P(t)] = \frac{\lambda}{s}$, where the expectation is with respect to the stationary Poisson distribution of $P(t)$. As a result, the total expected cost is
\begin{equation}
\mathbb{E}[C(P(t))] = \mathbb{E}[C(P(t))] = \sum_{i=0}^{\infty} q_i C(i)\,.
\label{eq:default}
\end{equation}
Given the cost function $C(\cdot)$, we can get the expression for the total expected cost.

\subsubsection{Variable power requirement per task}

The extension to different power requirements of tasks is done by reasoning as follows. Suppose that the power requirement of each task, $\hat{P}$ is a random variable that obeys a discrete probability distribution on values $\{p_1,\ldots,p_L\}$ with associated probabilities $w_1, \ldots, w_L$ (the case of continuous probability distribution of $\hat{P}$ is tackled similarly). Random variable $\hat{P}$ is taken to be independent from process $N(t)$. Let $\mathbb{E}[\hat{P}] = \sum_{k=1}^L p_k w_k$ be the expected value of power requirement. Power consumption at time $t$ is $P(t) = \hat{P} \cdot N(t)$, and the average power consumption at steady state is $\mathbb{E}[P(t)] = \lambda \mathbb{E}[\hat{P}]/s$.

This becomes obvious by the following analogy. For fixed, unit power requirements, $p_n = 1$, a demand request that arrives in infinitesimal time interval $[k\Delta, (k+1)\Delta]$ goes to one server in the $M/M/\infty$ system and is served; at that interval, the arrival rate is $\frac{1}{\Delta}$. If $p_n = n$, the situation is as if $n$ servers are occupied, or equivalently $n$ requests of unit power requirement appear in the same interval, and the arrival rate is $n \cdot \frac{1}{\Delta}$. Thus, an average power requirement $\mathbb{E}[\hat{P}]$ is equivalent to an average arrival rate $\lambda \mathbb{E}[\hat{P}]$ of requests with unit power requirement.

The total expected cost is found by taking expectation with respect to both the distribution of $N(t)$ and $\hat{P}$,
\begin{equation}
\mathbb{E}[\hat{P} \cdot C(N(t))] = \sum_{i=0}^{\infty} \sum_{k=1}^L q_i p_k C(i \cdot w_k)\,.
\label{eq:default2}
\end{equation}

The default policy described above activates each task upon arrival without taking into account system state information.
%The performance of this policy, given by (\ref{eq:default}) or (\ref{eq:default2}) is an upper bound on the performance of any scheduling policy.

\subsection{A Universal Lower Bound}

We now derive a lower bound on the performance of any scheduling policy in terms of total expected cost.

\begin{theorem}
The performance of any scheduling policy is at least $C\left(\lambda \mathbb{E}[\hat{P}] / s\right)$.
\end{theorem}

\begin{proof}
We use Jensen's inequality which says that for a random variable $X$ and convex function $C(\cdot)$, it is $\mathbb{E}[C(X)] \geq C(\mathbb{E}[X])$. Equality holds if and only if $X = \mathbb{E}[X]$, i.e when random variable $X$ is constant. Jensen's inequality in our case means
\begin{equation}
\mathbb{E}[C\big(P(t)\big)] \geq C\big(\mathbb{E}[P(t)]\big)\,.
\label{eq:ineq1}
\end{equation}
We now argue that this lower bound is universal for all scheduling policies. A scheduling policy essentially shifts arising power demand tasks in time. These time shifts alter instantaneous power consumption $P(t)$ and thus they can also change the steady-state distribution of $P(t)$. However, the average power consumption $\mathbb{E}[P(t)]$ in the system always remains the same.

To see this more clearly, consider the subsystem that includes only the power demands under service currently. The arrival rate at the subsystem is $\lambda \mathbb{E}[\hat{P}]$, and the time spent by a customer in the subsystem is $1/s$ regardless of the control policy. By using Little's theorem, we get that the average number of customers in the subsystem (which also denotes the average power consumption) is fixed, $\mathbb{E}[P(t)] = \lambda \mathbb{E}[\hat{P}]/s$, and the proof is completed.
\end{proof}

As will be shown in the sequel, this bound is asymptotically tight as the deadlines become larger. In other words, we will show that there exists a policy that is asymptotically optimal and achieves the bound.

\subsection{An Asymptotically Optimal Policy: Controlled Release}

Without loss of generality, assume unit power requirements, $p_n = 1$. Consider the following threshold based control policy. There exists a threshold $P_0$. Upon arrival of a new request at time $t$, the controller checks current power consumption $P(t)$. If $P(t) < P_0$, the demand request is activated, otherwise it is queued. Queued demands are activated either when their deadline expires or when the power consumption $P(t)$ drops below $P_0$. We refer to this policy as the \tit{Controlled Release (CR)} policy.

\begin{figure}[htb]
\begin{center}
\epsfig{figure = 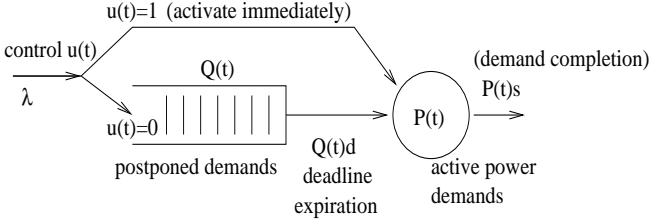, width=8.8cm,height=3cm}
\end{center}
\caption{Depiction of optimal threshold policies. The Threshold Postponement (TP) policy $\pi_b$ is depicted above. The Enhanced Threshold Postponement (ETP) policy $\pi_e$ follows the rationale depicted above, with rate $Q(t)d$ substituted by $Q(t)d + P(t)s \cdot \mathbf{1}[P(t) < P_e]$. The control $u(t) \in \{0,1\}$ is applied based on corresponding thresholds $P_b, P_e$ on power consumption $P(t)$.} \label{fig:policy}
\end{figure}

\begin{theorem}
The CR policy is asymptotically optimal in the sense that for optimized threshold, as the deadlines increase, its performance converges to the lower bound of all policies, $C\big(\mathbb{E}[P(t)]\big) = C\big(\frac{\lambda}{s}\big)$.
\end{theorem}

\begin{proof}
We provide a sketch of the proof. Consider an auxiliary system, $S_{\mbox{aux}}$ that is like the one described in the CR policy above, except that there are no deadline considerations. That is, in $S_{\mbox{aux}}$, upon arrival of a demand request at time $t$, the controller checks power consumption $P(t)$. If $P(t) < P_0$, the demand request is activated, otherwise it is queued. Queued demands are activated when the power consumption drops below $P_0$.

Clearly, in the auxiliary system, requests are queued only when the upper bound $P_0$ on power consumption is exceeded. Essentially $S_{\mbox{aux}}$ is equivalent to an $M/M/c$ queueing system, with $c = P_0$ ``servers'' \cite[Section 3.4]{bertsekas}. From Little's theorem, the average number of power demands in the system is $\lambda (\frac{1}{s} + W)$, where $W$ is the average waiting time of a request in the queue until it gets activated. Define the occupation rate per server, $\rho = \lambda /(cs)$. The average number of power demands in the system is written as $c \rho + \lambda W$. Note that term $c \rho$ denotes the expected number of busy servers at steady-state.

Now define a sequence of thresholds $P_0^n = \frac{\lambda}{s} + \epsilon_n$, $n=1,2,\ldots$, where $\epsilon_n$ is chosen so that $\lim_{n +\rightarrow \infty} \epsilon_n = 0$. Note that a sequence of occupation ratios $\rho_n$, $n=1,2,\ldots$, accordingly emerges, with $\rho_n = \lambda/(c_n s) = \lambda/(P_0^n s)$, and
\begin{equation}
\lim_{n \rightarrow +\infty} \rho_n = \lim_{n \rightarrow +\infty} \frac{\lambda}{s(\displaystyle \frac{\lambda}{s} + \epsilon_n)} = 1
\end{equation}
and therefore in the limit, the number of busy servers is constant, $\lambda/s$ with probability $1$. This implies that (\ref{eq:ineq1}) holds with equality and therefore the expected cost for the auxiliary system is $C\big(\frac{\lambda}{s}\big)$, which is precisely the universal lower bound derived above.

Consider now the original system with the CR policy. Queued requests are activated either when power consumption drops below $P_0$ or when the deadlines expire. The latter occurs with average deadline expiration rate $d$. As average deadline durations $1/d$ increase, the deadline expiration rate $d$ goes to $0$, and the original system tends to behave like the auxiliary system $S_{\mbox{aux}}$. Since the performance of CR policy converges to that of the auxiliary system as the deadlines increase, and the performance of the auxiliary system asymptotically achieves the lower bound above, it follows that the CR policy is also asymptotically optimal as deadlines increase.
\end{proof}

\subsection{Optimal Threshold Based Control Policies}

In this section we describe two power demand control policies that rely on instantaneous power consumption to make their decisions, yet their associated control spaces differ. We omit the proofs of optimality due to space limitations.

\subsubsection{Bi-modal control space}

First, we consider the class of bi-modal control policies for which the control space for each power demand task $n$ is bi-modal, namely $\mathcal{U}_b = \{0, D_n\}$. That is, each demand $n$ is either scheduled immediately upon arrival, or it is postponed to the end, such that it is completed precisely at the time when its deadline expires. Without loss of generality, we assume that  power requirements are fixed, $p_n = 1$.

We consider the following threshold policy $\pi_b$. At the time of power demand request arrival $t$, the controller makes the decision whether the demand will be served immediately or at the end of its deadline. If the total instantaneous power consumption $P(t)$ is less than a threshold $P_b$, the controller serves the power demand request immediately. Otherwise, if $P(t) > P_b$, it postpones the newly generated request to the end of its deadline. We call this policy, the \tit{Threshold Postponement (TP)} policy.

The system state at time $t$ is described by the pair of positive integers $\big(P(t),Q(t)\big)$ where $P(t)$ is the number of demands that consume power at time $t$ and $Q(t)$ is the number of postponed demands. Observe that there is an additional source of demand requests that enter power consumption with rate $Q(t) d$ where $d$ is the rate of deadline expiration.

Assuming that demand durations and deadlines are exponential and homogeneous and demand power level is fixed, $\big(P(t),Q(t)\big)$ is a controlled continuous time Markov chain. Define the control function $u(t)=1$ if newly arrived demands are activated immediately, and $u(t) = 0$ if they are postponed until their deadlines expire. The transitions that describe the continuous time evolution of the Markov chain are as follows. From state $(P(t),Q(t))$, there is transition to the state:
\begin{itemize}
\item $(P(t)+1, Q(t))$ with rate $\lambda u(t)$, which occurs when a new arriving demand is activated immediately.
\item $(P(t), Q(t) +1)$ with rate $\lambda (1-u(t))$, when a new demand is postponed and joins the queue of postponed demands.
\item $(P(t)-1,Q(t))$ with rate $P(t)s$, due to completion of active demands.
\item $(P(t)+1,Q(t)-1)$ with rate $Q(t)d$, due to expiration of deadlines of postponed demands.
\end{itemize}

When $P(t) < P_b$ then $u(t)=1$. Then, $P(t)$ varies with rate $\lambda + Q(t) d - P(t)s$ due to new requests, expirations of deadlines of postponed requests and completions of active demands. In the same case, $Q(t)$ decreases with rate $Q(t) d$. On the other hand, when $P(t) > P_b$ then $u(t)=0$; $P(t)$ varies with rate $Q(t) d - P(t) s$, while $Q(t)$ varies with rate $\lambda -Q(t) d$. The rationale of the TP policy is shown in Fig. \ref{fig:policy}.

\begin{theorem}
The policy that minimizes $\mathbb{E}[C\big(P(t)\big)]$ over all bi-modal control policies with control space $\mathcal{U}_b$ is of threshold type, where the threshold is a switching curve $P_b(Q)$ that is non-decreasing in terms of $Q$. For appropriately selected switching curve $P_b(Q)$, the TP policy above is optimal.
\end{theorem}

The proof is based on showing that the infinite horizon discounted-cost problem
\begin{equation}
\min \lim_{T \rightarrow \infty} \int_0^T \!\beta^t C\big(P(t)\big)\,dt
\end{equation}
with discount factor $\beta < 1$ admits a stationary optimal control policy. The long-run average cost problem is then treated as a limiting case of the discounted-cost problem as $\beta \rightarrow 1$ and has a stationary policy as well \cite{ross}.

Some intuition on the form of the switching curve could be obtained as follows. There must exist a value of $P(t)$, $P_b(Q)$, beyond which it is more probable to induce lower cost by serving a demand in the future than by serving it immediately with the current cost. From the transition rates above, observe that the likelihood of reducing $P(t)$ increases with increasing $P(t)$. Furthermore, the likelihood of reducing $P(t)$ goes down with increasing $Q(t)$, which means that it is more possible to increase $P(t)$ with increasing $Q(t)$. This seems to imply that  threshold $P_b(Q)$ is a non-decreasing function of $Q$.

\subsubsection{Enhanced control space}

Consider now the enhanced policy $\pi_e$. At the time of power demand request arrival $t$, the controller makes the decision whether the demand will be served immediately or at the end of its deadline. If the total instantaneous power consumption $P(t) \leq P_e$, the controller serves the power demand request immediately. Also, in this case, whenever an active power demand is completed, a postponed demand from the queued ones is activated. Otherwise, if $P(t) > P_e$, it postpones the newly generated request to the end of its deadline. Whenever the deadline of the demand expires, the demand is activated. This policy has the additional degree of freedom to schedule a demand after it is generated and before its deadline is expired. The control space for this policy is $\mathcal{U}_e = \{[a_n,D_n]\,\,\mbox{for}\,\,n=1,2,\ldots\}$, and clearly $\mathcal{U}_e \supseteq \mathcal{U}_b$. We call this policy, \tit{Enhanced Threshold Postponement (ETP)} policy.

The system state at time $t$ is again described by $(P(t),Q(t))$ where $P(t)$ is the number of demands that consume power at time $t$ and $Q(t)$ is the number of postponed demands. The control function is again defined to be $u(t)=1$ if newly arrived demands are activated immediately, and $u(t) = 0$ if they are postponed until their deadlines expire. The transitions from state $(P(t),Q(t))$ in the Markov chain are towards state:
\begin{itemize}
\item $(P(t)+1, Q(t))$ with rate $\lambda u(t)$, which occurs when a new arriving demand is activated immediately.
\item $(P(t), Q(t) +1)$ with rate $\lambda (1-u(t))$, when a new demand is postponed and joins the queue of postponed demands.
\item $(P(t)+1,Q(t)-1)$ with rate $Q(t)d$, due to expiration of deadlines of postponed demands.
\item $(P(t)-1,Q(t))$ with rate $P(t)s (1-u(t))$, due to completion of active demands, and no activation of queued demands.
\item $(P(t),Q(t)-1)$ with rate $P(t)s u(t)$, due to completion of active demands, and simultaneous activation of queued demands (that is why $P(t)$ does not change).
\end{itemize}

When $P(t) < P_b$ then $u(t)=1$. Then, $P(t)$ varies with rate $\lambda + Q(t) d - P(t)s$ due to new requests, expirations of deadlines of postponed requests and completions of active demands. In the same case, $Q(t)$ decreases with rate $Q(t) d$. On the other hand, when $P(t) > P_b$ then $u(t)=0$; $P(t)$ varies with rate $Q(t) d - P(t) s$, while $Q(t)$ varies with rate $\lambda -Q(t) d$. The rationale of the TP policy is shown in Fig. \ref{fig:policy}.

When $P(t) \leq P_e$ then $u(t)=1$. Then, $P(t)$ increases with rate $\lambda + Q(t) d$ due to the arriving request rate and the rate with which deadlines of postponed requests expire. Also $P(t)$ decreases with rate $P(t) s$ due to completion of active demands, but it also increases with rate $P(t)s$ since queued demands enter service whenever active ones are completed. When $P(t) \leq P_e$, $Q(t)$ decreases with rate $Q(t)d + P(t)s$. On the other hand, when $P(t) > P_e$ then $u(t)=0$; $P(t)$ varies with rate $Q(t)d - P(t)s$, while $Q(t)$ varies with rate $\lambda -Q(t) d$. The ETP policy $\pi_e$ follows the rationale depicted in Fig. \ref{fig:policy}, with rate $Q(t)d$ substituted by $Q(t)d + P(t)s \cdot \mathbf{1}(P(t) < P_e)$, where $\mathbf{1}(\cdot)$ is the indicator function.

\begin{theorem}
The policy that minimizes $\mathbb{E}[C\big(P(t)\big)]$ over all control policies with control space $\mathcal{U}_e$ is of threshold type. For appropriately selected threshold $P_e$, the ETP policy is optimal.
\end{theorem}

Here, the threshold $P_e$ does not depend on $Q(t)$, since $P(t)$ is fed with queued demands when a demand is completed, and therefore it will remain approximately around a fixed threshold as long as $Q(t)$ is not empty.

\section{Discussion} \label{sec:5}

In this work, we took a first step towards bringing control and optimization theory in the context of smart power grid. We focused on a scenario where control of consumer appliances is fully delegated to the grid operator, and we studied the fundamental problem of smoothing the power demand profile so as to minimize the grid operational cost over some time horizon and promote efficient energy management. This problem is envisioned to be a central one with smart grid-enabled appliances and two-way communications between the provider and consumers. First, we studied the off-line version of the scheduling problem. The optimal solution was derived for elastic demands that allow preemptive scheduling, while for inelastic demands that require non-preemptive scheduling the problem is NP-Hard. In light of a dynamically evolving system and the need for online scheduling decisions, we studied long-term expected cost through a stochastic model. Our main result is a threshold scheduling policy with a threshold on instantaneous power consumption, which is asymptotically optimal in the sense of achieving a universal lower performance bound as deadlines increase. We have also proposed two control instances with different control spaces. In the first one, the controller may choose to serve a new demand request upon arrival or to postpone it to the end of its deadline. The second one has the additional option to activate one of the postponed demands when an active demand terminates. For both instances, the optimal policies are threshold-based.

There exist many issues for investigation. For the threshold based policies that we described, an elaborate study and derivation of the structure of the policies and threshold values would enhance our study. Some input from real-life power grid systems in terms of the operational cost and power demand statistics would positively modulate the process of explicit computation of thresholds.

For the scenario envisioned in this paper, the incorporation in the model of different classes of power demand tasks with different inherent constraints is of great interest. Different classes of tasks, some of which were captured by the current formulation, are as follows:
\be
\item Demands that may have fixed power requirement and zero time tolerance in scheduling, e.g. lights.
\item Demands that have fixed power requirements and there exists some flexibility in scheduling within a certain time window, e.g. washing machine or dishwasher.
\item Demands that have flexibility both on the power demand and the duration. Some of these may need to be periodically turned on and off by the operator, like the air conditioning.
\item \tit{Special} types of demands. For example, in the task of charging electric vehicles, there exist constraints on the total amount of energy needed to charge the battery and on the time interval by which charging needs to be completed. Charging may take place at  nonconsecutive time intervals and with adaptable charging rate. The latter results in a flexibility in tuning instantaneous power demand.
\ee
Especially the possibility of controlling the power consumption level of appliances in addition to time scheduling adds a new dimension to the problem. Such scenarios have already started finding their way in instances where the consumption level of consumer A/C is controlled by the operator. The derivation of optimal control policies in this context is an interesting issue.

In this work, we assumed that the provider has full control over consumer appliances, and these always comply to the dictated schedule. A lot of other scenarios could be envisioned. For instance, some freedom may be granted to the consumer to select whether the announced schedule by the provider will be admitted or not. Some incentives from the provider side could also be considered in that case, like reduced prices if users comply to the schedule. If continuous feedback on instantaneous price per unit of power demand is provided by the operator, the user would need to decide whether to activate the demand immediately and pay the instantaneous price, or postpone the demand for a later time, if such an option exists, with the hope that the price becomes lower. Another possibility in that case could be that each consumer makes its proposition to the provider in terms of defining its time flexibility in scheduling according to the announced price. Each of the scenarios above gives rise to interesting mathematical models of interaction that warrant investigation.

%If power consumers are also power micro-generators, joint orchestration of power demand profiles of all consumers is needed. In that case, the possibility that micro-generators may offer their excess load to the grid at certain prices needs to be considered. Another possibility that alters the model significantly is to introduce energy storage capabilities at the consumer side.

%\subsubsection{Controlled release policy}
%Newly generated demands are queued at generation time.
%When the deadline of a demand expires then the demand is scheduled automatically.
%When a new demand is generated or a demand under service terminates then one of the queued %demands may be scheduled for service. The combination of earliest deadline first priority rule %for the queued demands, with a  threshold rule, should work well in this case.
%Computation of the optimal threshold is an issue here as well.

%Whenever a demand can be scheduled without violating the upper bound, it is done so. If the %upper bound is violated, delay the demand if there is slack time. When there is space to
%schedule the demand do so with the EDF rule.

%Another special case where the optimal policy might be characterizable: service time and
%deadlines deterministic, demand power level fixed.

%Optimize with respect to the selection of the threshold. Deadlines known or unknown.

% IMPORTANT: Can I use the outcome of the offline problem (waterfilling-load balancing) to derive a threshold for the stochastic problem?


\begin{thebibliography}{1}

\bibitem{moslehi} K. Moslehi and R. Kumar, ``Smart Grid: A Reliability Perspective'', \tit{Proc. IEEE PES Conference on Innovative Smart Grid Technologies,} 2010.

\bibitem{lui} T.J. Lui, W. Stirling and H.O. Marcy, ``Get Smart'', \tit{IEEE Power and Energy Mag.,} vol.8, no.3, pp.66-78, May/June 2010.

\bibitem{link1} http://www.smartgridnews.com/artman/publish/Technologies\_Demand\_ Response/

\bibitem{hamilton} K. Hamilton and N. Gulhar, ``Taking Demand Response to the Next Level'', \tit{IEEE Power and Energy Mag.,} vol.8, no.3, pp.60-65, May / June 2010.

\bibitem{link2} GridWise Initiative. http://www.gridwise.org/.

\bibitem{link3} Peaksaver Program. https://www.peaksaver.com/peaksaver\_THESL.html.

\bibitem{link4} Lockheed Martin SEELoad\texttrademark Solution: http://www.lockheedmartin. com/data/assets/isgs/documents/EnergySolutionsDataSheet.pdf.

\bibitem{link5} Managing Energy with Swarm Logic, \tit{MIT Technology Review,} online: http://www.technologyreview.com/energy/22066/.

\bibitem{li} H. Li and R.C. Qiu, ``Need-based Communication for Smart Grid: When to Inquire Power Price?'', \tit{http://arxiv.org/abs/1003.2138}.

\bibitem{handbook} J. Y.-T. Leung (\tit{Ed.}), \tit{Handbook of Scheduling: Algorithms, Models and Performance Analysis,} Chapman and Hall / CRC, 2004.

\bibitem{panwar} S. Panwar, D. Towsley, and J. Wolf, ``Optimal scheduling policies for a class of queue with customer deadlines to the beginning of services'', \tit{J. Assoc. Comput. Mach.,} vol.35, no.4, pp.832–844, 1988.

\bibitem{keslassy} I. Keslassy, M. Kodialam and T.V. Lakshman, ``Faster Algorithms for Minimum Energy Scheduling of Data Transmissions'', \tit{in Proc. WiOpt,} 2003.

\bibitem{neely} W. Chen, M.J. Neely and U.Mitra, ``Energy Efficient Scheduling with Individual Packet Delay Constraints: Offline and Online Results'', \tit{in Proc. INFOCOM,} 2007.

\bibitem{fu1} A. Fu, E. Modiano and J.N. Tsitsiklis, ``Optimal Transmission Scheduling over a Fading Channel with Energy and Deadline Constraints'', \tit{IEEE Trans. on Wireless Comm.,} vol.5, no.3, pp.630-641, March 2006.

\bibitem{tarello} A. Tarello, J. Sun, M. Zafer and E. Modiano, ``Minimum Energy Transmission Scheduling subject to Deadline Constraints'', \tit{in Proc. WiOpt,} 2005.

\bibitem{zafer} M.A. Zafer and E. Modiano, ``A Calculus Approach to Minimum Energy Transmission Policies with Quality of Service Guarantees'', \tit{IEEE/ACM Trans. Networking,} vol.17, no.3, pp.898-911, June 2009.

\bibitem{kumar1} I.-H. Hou and P.R. Kumar, ``Utility Maximization for Delay Constrained QoS in Wireless'', \tit{in Proc. INFOCOM,} 2010.

\bibitem{kumar2} I.-H. Hou and P.R. Kumar, ``Scheduling Heterogeneous Real-time Traffic over Fading Wireless Channels'', \tit{in Proc. INFOCOM,} 2010.

\bibitem{hajek} B. Hajek, ``Performance of Global Load Balancing by Local Adjustment'', \tit{IEEE Trans. Inf. Theory,} vol.36, no.6, pp.1398-1414, Nov. 1990.

\bibitem{garey} M.R. Garey and D.S. Johnson, \tit{Computers and Intractability: A guide to the Theory of NP-Completeness,} Bell Tel. Lab., 1979.

\bibitem{bertsekas} D. Bertsekas and R. Gallager, \tit{Data Networks,} Prentice-Hall, 2nd Ed., 1987.

\bibitem{ross} S. M. Ross, \tit{Introduction to stochastic dynamic programming,}  New York Acad. Press, 1983.
\end{thebibliography}
\end{document}